# Application Security framework for Mobile App Development in Enterprise setup

**\*Subhamoy Chakraborti**
Magma Fincorp Limited, India
Email: chakrabortisubhamoy@gmail.com
**D. P. Acharjya**
School of Computing Science and Engineering, VIT University, Vellore, India
E mail: dpacharjya@gmail.com
**Sugata Sanyal**
Corporate Technology Office, Tata Consultancy Services, Mumbai, India
Email: sugata.sanyal@tcs.com
\*Corresponding author

**ABSTRACT**
Enterprise Mobility has been increasing the reach over the years. Initially Mobile devices were adopted as consumer devices. However, the enterprises world over have rightly taken the leap and started using the ubiquitous technology for managing its employees as well as to reach out to the customers. While the Mobile ecosystem has been evolving over the years, the increased exposure of mobility in Enterprise framework have caused major focus on the security aspects of it. While a significant focus have been put on network security, this paper discusses on the approach that can be taken at Mobile application layer, which would reduce the risk to the enterprises.

## 1. INTRODUCTION
Enterprise Mobility has been a buzzword for quite some time. While Mobility has increased its reach rapidly as a consumer electronic device, smart enterprises have been using the technology to its benefit. While the major focus areas have been in device, application & network side, the importance of security is also emerging now. CIOs have been approaching the Enterprise Mobility initiatives mostly from Application side. Due to its lack of focus, this paper attempts to discuss a Mobile App Security Framework. This can help any Enterprise which is embarking on its Enterprise Mobility journey to reduce the risks related to Enterprise Data, prospective customers & demographic details.

## 2. OVERVIEW
Mobility security standards in Enterprise setup requires a tad different approach compared to consumer apps or Gaming apps. The major difference of Enterprise Apps with the Gaming or consumer apps is that the former has customer data and enterprise business logic at its core. This is very critical in Banking & Financial Institutions [1].

In current days, the most popular Mobile operating systems have been Android and Apple iOS in consumer space. Amongst them, Android operating system has majority of the market share [2]. The consumer market trend has been visible in Enterprise scenario also [16].

The discussion in this paper revolves around the Security framework for Mobile Application in general and Android operating system [3], [4] in specific where specific references are required. The potential attack type in general to Mobile devices and the goal & vector can be of wide variety [9]. But this paper puts more focus on the impact to vulnerabilities in Enterprise Mobile Apps.

In Section 3, the paper defines 4 broad areas of security concerns in Enterprise Mobile application context. In Section 4, the paper suggests technical approaches to minimize the risks aforementioned from application development point of view. Section 5 concludes on the approach towards implementing a secure Mobile application development framework for any enterprise.

## 3. AREAS OF CONCERN
In this paper, we have classified the security concerns around Mobile App development into 4 broad areas.

(a) Data protection
(b) Intellectual property protection
(c) Secure authentication
(d) Code vulnerability

We would cover each of these areas in more detail in the following sub-sections.

## 3.1 DATA PROTECTION

Mobile devices inherently pose a challenge regarding the data vulnerability due to its ubiquitous nature and dependence on transmission of data over the air. In Enterprise scenario, the risk is more complex than other categories of apps like gaming app as the data in Enterprise app may consist of financial details or demographics of a customer. Losing this data, in transit or while at rest, may have direct or indirect impact to company's revenue and / or reputation. Losing data about prospective customers may directly result in loss of business. Losing demographics details or sensitive information about customers may result in legal issues as well. Mobile Application can be developed either as a native application tightly stitched to the operating system or the App can be developed as Web Apps [6]. In either of these approaches, Android provides a permission based access mechanism to the apps [8]. However these permissions do not dictate data policies to be adopted in the apps. Based on the type of vulnerabilities, we approach the Data Protection issue in 4 parts.

(a) Local data store: Most of the mobile operating systems contain a lightweight database as part of the stack. Applications generally store the persistent data in this lightweight database. If data is stored as clear text in this local database, it can pose a serious security threat. Access to the database may result in loss of data. Also attackers may tamper this data causing malfunctioning of the application itself.

(b) Cache usage: Cache is an important way to enhance the user experience in Mobile operating system [22]. Cache can be at multiple levels of the architecture. A major usage of cache is in web applications, where it is used to store data across sessions to give a consistent user experience. However if critical information is stored in the cache in unsecure way like plain text, other phishing applications may attempt to exploit the cache causing loss or tampering of data.

(c) Data sharing: Data loss can also happen due to not partitioning the application properly. Android imposes each application to run in a separate instance of its proprietary Virtual machine [7], which isolates the applications from each other. However data vulnerability can still exist if the application stores its data in removable storage medium in unencrypted format. In such cases, though the application execution would be contained within the context of the particular virtual machine, the app can still become vulnerable to attacks through file system.

(d) Data on transit: The other major risk to focus is data on transit. Securing the data transmission over the network has been of academic and industrial focus [19], [20], and [21]. The major dependence of protecting data on transit is on the underlying network security. Since tapping into wireless data stream is mostly a trivial problem, forming the data packet securely to be sent over network requires more attention. This would reduce the risk of data loss even in cases of data packet capture from the wireless network by rogue elements. The loss can happen during data requests over TCP/IP as well as over insecure SMS protocol used for application to application messaging.

All the scenarios mentioned in this section might cause loss of data or tampering of data. As mentioned, the data loss can result in financial or reputation loss for the enterprise and hence critical to address properly.

## 3.2 INTELLECTUAL PROPERTY PROTECTION

The second major cause of loss of information is through unauthorized access to the application code base. Apple iOS uses .IPA (iPhone Application) extension while Android uses .APK (Android Application Package) to distribute the application binary in the App Store or Play store. Since the binaries are easily available, any attacker may like to do reverse engineering of the application binary to obtain the source code.

(a) Reverse Engineering: Android applications are written in Java like language which is compiled to generate byte code compatible with the proprietary Virtual machine. Though it is non-trivial, but it is possible to reconstruct the source code from the application binary, i.e. apk files by using certain decompilers [5]. Similar approach can be used in other Mobile platforms. This may result in exposure of critical information about the enterprise embedded within the code.

(b) Critical information hardcoded: Also code access may lead to revealing crypto keys and user credentials. If the enterprise app deals with payment, this may also lead to exposing the payment access details leading to monetary impact. Leakage of all these critical information to any attacker can put the enterprise at major risk.

### 3.3 SECURE AUTHENTICATION
In the commonly adopted Enterprise Mobile App architecture, the apps connect with a middleware application which in turn connects with the core system at the backend [1]. Authentication of the user at the mobile end needs to be planned keeping security in mind. The possible risks may arise due to improper handling of data exchange between the Mobile app and the middleware. Also this may be caused due to the quality of the password string.
   (a) Session management: Sessions are setup between the Mobile App & the middleware application to handshake status, information and master data. Application Sessions need to be handled carefully to avoid session hijacking or session fixation attacks.
   (b) Password management: Authentication to the app needs to be made secure in such a way that even if the credentials are lost, still this would not result in adversary having access to the application. Setting up the password with desirable complexity is the first step. Possible causes of attack may include predictable password. For a group of employees, this would result in availability of access to all the devices by anyone.

### 3.4 CODE VULNERABILITY
Vulnerability in the application design and coding may lead to massive security issues.
   (a) Validation: The mobile applications vastly uses scripting languages for developing the front end. This is highly risky as the scripting language can be modified. The data entry through the front end goes through a front end validation. However front end validation is not secure enough as scripting can be changed through similar-looking UI to send malicious code. Through this method, SQL command can also be passed through data entry form to retrieve application details or in worst case scenario, alter the application database at the backend. SQL injection attacks are very much possible in Mobile applications, which can be a major threat to Enterprise mobility applications.
   (b) Exception Handling: Another possible source of leakage of information is improper exception handling of the code. Raw exception dump can lead to exposing business logic to the user. This can help the attacker identify possible areas to attack the application as the possible business logic can be deciphered from a stack trace of the application.
   (c) Other source code: Many times application source rely on 3$^{rd}$ party libraries and other source codes. Though development team focus on maintaining security standards in own code, often vulnerability may arise due to these 3$^{rd}$ party code. Attackers can utilize broken cryptographic algorithm or commonly used libraries to gain access to the application, even if the application may not have any vulnerability on its own.

## 4. PROPOSED APPROACHES OF MITIGATING THE RISKS
In the previous section, we discussed about common concern areas from Mobile Application Security standpoint. With a systemic approach, it is possible to mitigate these risks to a large extent and minimize the risks.

### 4.1 DATA PROTECTION
We mentioned about the possible issues around securing data in the ubiquitous devices in section 3.1. In this section we would discuss about approaches which can be used to reduce the risk in this area.
   (a) Local data store: Application stores certain critical data in the local data store, generally in a lightweight database like SQLite. Application design should be made in such a way that the size and duration of data storage can be minimized. Data audit needs to be done to check the criticality of the data that remains in the device even for a short span of time. However there would be certain data, for example user role and access related masters, which are required for offline access to the application. This data must be encrypted in the data store. In the untoward scenario of the device being in physical custody of an adversary, application design decision needs to be made to minimize the ease of reading the local store data. Also application level granular check needs to be put in place to ensure that local device database modification can only happen through the application code. That would avoid data modification by any other source causing malfunctioning of the app.
   (b) Cache usage: Web applications need to store certain footprint in the cache to enhance the

performance of the application or to identify the device in the subsequent operations. However care needs to be taken that this cache doesn't contain any critical information about the customer or the organization. This would reduce the access of customer data by other app or browser based services.

(c) Data sharing: Applications should clearly partition the data within its boundary. The data can also include images of documents especially in Financial Institutions as that help in reducing the decision time for any loan application [13] [14] 15]. The application footprint should be partitioned into areas based on sensitivity and vulnerability of the data. Authentication mechanism must be put in place to restrict movement of data from secure area to unsecure area.

(d) Data on transit: Application should encrypt the message to be sent over the air. Also the data communication channel should be secured via HTTPS instead of plain HTTP. SSL must be implemented to send any application data including session data, sensitive information and tokens. Data exchange that happens over SMS channel must also be encrypted. All the data transactions happening over SMS protocol should be over Secure SMS protocol, example Secure SMPP. Key exchange based SMS protocol can help in this context [23].

## 4.2 INTELLECTUAL PROPERTY PROTECTION

We discussed about concerns around secure application code in section 3.2. This section covers approaches in mitigating those risks.

(a) Reverse Engineering: Android compiler creates byte code out of the source code. This byte code retains lots of debugging information including file name, class details, line number etc. Using this, a decompiler can reverse engineer the byte code into high level code, which may not be a desirable scenario for any enterprise as discussed in section 3.2. To avoid loss of leakage of the proprietary information via reverse engineering of the application binaries, obfuscation must be done. Through Obfuscation, the debugging information can be removed as well as the object names within the byte code can be mangled or replaced by meaningless entities, without hampering the way the application works. This helps hide the business logic of the application even if the adversary generates the code by reverse engineering of the application binary.

(b) Critical information hardcoded: It must be ensured that no critical information is getting hard-coded within the application code, including but not limited to crypto keys, user credentials, and other sensitive information like debit and credit card details. If the application is required to store crypto keys, it must ensure that the keys are protected in a key store instead of keeping it as part of application code.

## 4.3 SECURE AUTHENTICATION

Secure Authentication and related challenges were covered in section 3.3. This section discusses about technical approaches to mitigate risks on that front.

(a) Session management: To avoid eavesdropping and subsequent session replay kind of attack, session ID must be used for any transaction between the Mobile application and the middleware. Also the session id may be appended with additional unique information that identifies the device or user so that any unauthorized device cannot use the same password as pose as an authentic user to the application server. Possible ways of achieving the same can be by appending the session ID with device IMEI or MSISDN etc.

(b) Password management: While password management itself is an important topic, this paper discusses the important steps that a Mobile application must cater to while being deployed in Enterprise setup. The first level of vulnerability may arise due to availability of the password with unauthorized person. The application should depend not only on the password but rather should adopt multi factor authentication. Mobile devices are used to a large extent for securing password in untrusted computing end points [24] by setting up two factor authentication. However the challenge of building two factor authentication for access to mobile application is to find another trustworthy mobile device or a different mechanism, instead of sending the details to the same device. Password aging principle must be put in place to force the user to change the password on predefined intervals, thus reducing the risk. Also the application should retain the password history so that the user is not allowed to reset the password to the last 5 passwords. Algorithm can be built to detect dictionary words and stop the user from setting them as password as these are predictable by any attacker. Password complexity should be defined at the organization level to ensure minimum complexity is present instead of too simple passwords.

**4.4 CODE VULNERABILITY**

Code vulnerability poses a major threat as detailed in section 3.4. This section talks about approaches to protect against code vulnerability.

(a) Validation: To reduce the time to market, scripting languages have been majorly used in Mobile applications. While this gives a lot of flexibility and a good turnaround time, this also exposes the application to possible tweaks of the script at the front end and sending non-validated data to the server. Essentially the application may break due to this bypassing of validations at the front end. To avoid this, it is absolutely necessary to have replication of the front end validations at the server end also. This would block possible attacks through bypassing the front end scripts. Also server end validation would reduce the risk of SQL injection attacks through vulnerable UI design.

(b) Exception Handling: While it is important to capture the stack trace and make it available to the development team for analysis of possible issues in the application, it is also necessary to avoid showing the stack trace to the end user. Proper exception handling with customized messaging not only creates a better interactive app, but it also reduces the security exposure.

(c) Other source code: While using any source library, it is advised to enlist the deprecated APIs. While most of the platforms publishes the list of deprecated APIs, pre-processor based approach can be considered for other platforms where the deprecated APIs are not marked clearly.

**5. CONCLUSION**

Mobile operating systems, like Android, provide a system security model [10] [11] as part of its stack. This prevents many risks by virtue of Operating System level controls. However further planning needs to be done to take care of issues like data loss, Intellectual Property violation etc. We have discussed about specific issues and their possible solutions in the Enterprise Mobile App development context. Considering the application framework that has been mentioned in the paper, it is important for any organization embarking into Enterprise Mobility journey to have a clearly defined coding standard [12]. This standardization of approach towards coding standard and compliance to the same can help mitigate the security related issues in Mobile applications to a large extent and make the Enterprise Mobility endeavor successful.

# Biographies and Photographs

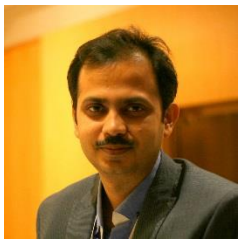

**Subhamoy Chakraborti** is working as General Manager at Magma Fincorp Limited, India. He has 12 years of experience in Technology Management. He has worked with Oracle India on ERP product development. He has also worked for various Tier 1 OEMs like Motorola, Toshiba, Fujitsu, Microsoft, Cisco, INQ in delivering phone programs while working for Product Engineering Services group in Wipro Technologies. He has keen interest in Data Science, Mobility and Cloud. His current focus includes building Enterprise Mobility ecosystem, Data driven decision making and Cloud Infrastructure. He has been working on defining the IT strategy and roadmap for his current organization with the CIO. Subhamoy participates in various campus focused programs and frequently contributes in CIO facing journals.

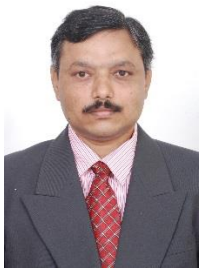

**D. P. Acharjya** received his M. Sc. from NIT, Rourkela, India; M. Tech. in computer science from Utkal University, India and obtained his Ph. D. in computer science from Berhampur University, India. He has been awarded with Gold Medal in M. Sc. He is presently working as a Professor in the School of Computing Sciences and Engineering at VIT University, India. He has authored more than 50 international, national papers, book chapters and four books entitled Fundamental Approach to Discrete Mathematics, Computer Based on Mathematics, Theory of Computation, and Rough Set in Knowledge Representation and Granular Computing to his credit. In addition, he has edited two books with IGI Global and one book with Springer, USA. His research interest includes rough sets, knowledge representation, machine learning and business intelligence. Dr. Acharjya is associated with many professional bodies CSI, ISTE, IMS, AMTI, ISIAM, OITS, IACSIT, CSTA and IAENG. In addition, he was founder secretary of OITS Rourkela chapter.

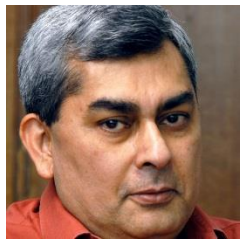

**Sugata Sanyal** is presently acting as a Research Advisor to the Corporate Technology Office, Tata Consultancy Services, India. He was with the Tata Institute of Fundamental Research till July, 2012. Prof. Sanyal is a: Distinguished Scientific Consultant to the International Research Group: Study of Intelligence of Biological and Artificial Complex System, Bucharest, Romania; Member, "Brain Trust," an advisory group to faculty members at the School of Computing and Informatics, University of Louisiana at Lafayette's Ray P. Authement College of Sciences, USA; an honorary professor in IIT Guwahati and Member, Senate, Indian Institute of Guwahati, India. Prof. Sanyal has published many research papers in International Journals and in International Conferences worldwide: topics ranging from network security to intrusion detection system and more.